\documentclass[apj]{emulateapj}
\usepackage{times}
\usepackage{graphics,epsf}
\usepackage{amsmath}                
\usepackage{amsfonts}               
\usepackage{amssymb}                
\usepackage{epsfig}                 
\usepackage{graphicx}
\usepackage{url}
\usepackage{tabularx}
\usepackage{booktabs}
\usepackage{hyperref}
\def \kms{\rm{km}$\rm{s}^{-1}$}

\def \s{~\rm{s}}

\def \km{~\rm{km}}
\def \kms{~\rm{km}~{\rm s}^{-1}}

\def \erg{~\rm{erg}}
\def \foe{~\rm{foe}}

\begin{document}

\title{Implications of turbulence for jets in core-collapse supernova explosions}

\author{Avishai Gilkis}
\author{Noam Soker}
\affil{Department of Physics, Technion -- Israel
Institute of Technology, Haifa 32000, Israel;
agilkis@tx.technion.ac.il; soker@physics.technion.ac.il}

\begin{abstract}
We show that turbulence in core collapse supernovae (CCSNe) which
has been shown recently to ease shock revival might also lead to
the formation of intermittent thick accretion disks, or accretion
belts, around the newly born neutron star (NS). The accretion
morphology is such that two low density funnels are formed along
the polar directions. The disks then are likely to launch jets
with a varying axis direction, i.e., jittering-jets, through the
two opposite funnels. The energy contribution of jets in this {\it
jittering jets mechanism} might result in an explosion energy of
$E_{\rm exp} \ga 10^{51} \erg$, even without reviving the stalled
shock. We strengthen the jittering jets mechanism as a possible
explosion mechanism of CCSNe.

\end{abstract}

\section{INTRODUCTION}
\label{sec:introduction}

Massive stellar cores undergo catastrophic collapse as their
final stage of evolution - this collapse is hypothesized to
result in energetic, luminous explosions termed core-collapse supernovae (CCSNe).
Of the several proposed theoretical explanations for the
explosion mechanism (see \citealt{Janka2012} for a review),
the most prominent are neutrino-driven explosions \citep{Colgate1966}
and jet-driven explosions (e.g. \citealt{LeBlanc1970,Khokhlov1999,Lazzati2012}).
The most modern neutrino-driven model is the
delayed-neutrino mechanism \citep{bethe1985, Wilson1985},
while jet-driven models have reincarnated in the jittering-jets mechanism
\citep{Papish2011, Papish2012b, PapishSoker2014a, PapishSoker2014b}.

In the delayed-neutrino mechanism,
neutrinos that are emitted by the newly formed neutron star (NS)
within a period of $t \approx 1 \s$
after the core bounce heat the region below the stalled shock,
at $r \approx 100-200 \km$ from the NS.
It has been suggested that subsequent neutrino-heating of the gain region will revive the stalled shock,
thereby exploding the star with the observed energy of $E_{\rm exp} \ga 1 \foe$,
where $1 \foe \equiv 10^{51} \erg$.

Widely varying outcomes have emerged from increasingly sophisticated multidimensional simulations
of the delayed-neutrino mechanism (e.g.,
\citealt{bethe1985,Burrows1985,Burrows1995,Fryer2002,Buras2003,
Ott2008,Marek2009,Nordhaus2010,Brandt2011,Hanke2012,Kuroda2012,
Hanke2012,Mueller2012,Bruenn2013,Mezzacappaetal2014,MuellerJanka2014a}).
Many of these failed to revive the stalled shock while others
produced tepid explosions with energies less than $1 \foe$.
In spherically symmetric calculations (1D),
the vast majority of progenitors cannot even explode
\citep{Burrows1995,Rampp2000,Mezzacappa2001,Liebend2005}.
The explosion of the $8.8 M_{\odot}$ progenitor of \citet{Nomoto1988}
in a 1D study with an energy of $\sim 3 \times10^{49} \erg$
is attributed to neutrino-driven wind \citep{Kitaura2006}.

In recent years, the standing accretion-shock instability (SASI;
e.g., \citealt{BlondinMezzacappa2003, BlondinMezzacappa2007,
Fernandez2010}) that appears in many two-dimensional axisymmetric calculations
\citep{Burrows1995,Janka1996,Buras2006a,Buras2006b,Ott2008,Marek2009}
has been studied as a possible driving force for stellar
explosions within the delayed-neutrino mechanism
(\citealt{Rantsiouetal2011} further suggested the spiral mode of
the SASI as the source of pulsar angular momentum). However,
recent three-dimensional studies gave mixed results
\citep{Nordhaus2010,Hanke2012,Hanke2013, Couch2013,
Dolence2013,Janka2013,CouchOConnor2014,Mezzacappaetal2014,Takiwakietal2013}.
While \cite{Nordhaus2010} and \cite{Dolence2013} found it easier
to revive the stalled shock in 3D simulations, most studies have
found that explosions are harder to achieve in 3D than 2D
\citep{Hanke2012,Couch2013,Janka2013, 
Hanke2013, CouchOConnor2014, Takiwakietal2013}.
Most striking is the comparison of
the 2D and 3D results of the Oak Ridge group. In their 2D
simulations \cite{Bruenn2013, Bruennetal2014} successfully revived
the shock with explosion energy estimates of approximately
$0.1-0.8 \foe$. However, in the newer 3D case presented by
\cite{Mezzacappaetal2014} the shock radius position is similar to
their results of 1D simulations where no explosion had been
obtained. A summary of some of these studies and an account of the
seemingly successful explosion of \cite{Bruenn2013} are given by
\citet{Papishetal2015}.

Even if the simulations overcome the problem of shock revival, in
most cases of unscaled simulations the explosion energy is lower
than required -- less than $1\foe$. \cite{Papish2012a} and
\cite{Papishetal2015} argued that there is a generic problem of
the delayed-neutrino mechanism that prevents it from exploding the
star with energies above $5 \times 10^{50} \erg$, and in most
cases much lower.

Recently \cite{CouchOtt2013}, \cite{CouchOtt2014}, and
\cite{MuellerJanka2014b} argued that the effective turbulent ram
pressure exerted on the stalled shock allows shock revival with
less neutrino heating than 1D models.
However, \cite{Abdikamalovetal2014}
found that increasing the numerical
resolution allows a cascade of turbulent energy to smaller scales,
and the shock revival becomes harder to achieve at high numerical resolution.
We here nonetheless study the implication of the turbulence on the
stochastic accretion of angular momentum onto the newly formed NS.
In section \ref{sec:oneelem} we show implications of accretion
of material from a convective region of the progenitor star
for formation of intermittent thick disks around the NS,
and in section \ref{sec:multielem} we discuss the
implications of accretion of many convective elements simultaneously.
In section \ref{sec:summary} we briefly discuss
the stochastic angular momentum in the post-bounce turbulent core,
and summarize.

\section{ACCRETION OF ONE CONVECTIVE ELEMENT}
\label{sec:oneelem}

In this section we present the method of calculation by studying one convective cell.
Here, and in the next section, we consider scenarios where accretion takes place through a disk.
The disk is optically thin to neutrinos, and optically thick to photons.
``Thin'' or ``thick'' disks (geometrically) are defined according to the ratio of the vertical disk height $H$
to the radial distance in the equatorial plane $r_\mathrm{e}$. 
If $H \ll r_\mathrm{e}$ then the disk is geometrically thin.
If $H \approx r_\mathrm{e}$ then the disk is geometrically thick.
If $H > r_\mathrm{e}$ and $r_\mathrm{e}$ is not much larger than the radius of the newly formed NS,
we term the disk a {\it belt}.

\subsection{Geometrically thin accretion disk}
\label{subsec:thindisk}

To demonstrate that the turbulent convection required to revive
the stalled shock can lead to intermittent disk formation we
consider a progenitor with an initial main sequence mass of
$M_\mathrm{ZAMS}=15M_{\odot}$ and a metallicity of $Z=0.014$. We
evolve the star using version 5819 of the Modules for Experiments
in Stellar Astrophysics (MESA; \citealt{Paxton2011,Paxton2013}).
Just before core collapse the velocity of convection in the
silicon layer, given by the mixing-length theory (MLT) employed by
MESA, has a Mach number of $\mathcal{M}_{c} \approx 0.01$.
However, some studies of realistic hydrodynamical simulations of
convection in stellar interiors show higher convective velocities
of $\mathcal{M}_{c} \approx 0.1-0.2$ \citep{Bazan1998,Asida2000}.
While recent studies
\citep{CouchOtt2013,CouchOtt2014,MuellerJanka2014b} have shown
that initial conditions motivated by these results alleviate the
required neutrino energy for a shock revival in the
delayed-neutrino mechanism, we focus on the implications for
stochastic angular momentum in the collapsing material, and
subsequently the possible formation of accretion disks and jets.
We emphasize that the convective velocity assumed here is not
what we take from numerical simulations of pre-explosion stellar
models, but rather the convective velocity used by
\cite{CouchOtt2013}, \cite{CouchOtt2014}, and
\cite{MuellerJanka2014b}. This is because our goal is to examine
the implications of the convection assumed by these  authors to
the formation of accretion disk or belts around the newly born NS.

Similarly to \cite{GilkisSoker2014}, where the details of the
calculations can be found, we calculate the variance of the
specific angular momentum. We assume a random velocity
$\overrightarrow{v}=v_{c}\left(\sin\theta\cos\varphi,\sin\theta\sin\varphi,\cos\theta\right)$,
where $v_{c}$ is the convective speed, with a uniform probability
density in $\theta$ and $\varphi$ (the angles relative to the
$z$-axis and $x$-axis, respectively - although the choice of axes
is inconsequential). The expectation value for the specific
angular momentum along a specific direction, here taken to be the
$z$ axis, is zero, while the variance is
\begin{multline}
\mathrm{Var}(j_{z})=\left\langle j_{z}^{2}\right\rangle =(v_{c}r_{l})^{2}\frac{\int\left[\left(\hat{r_l}\times\hat{v}(\Omega)\right)\cdot\hat{z}\right]^{2}d\Omega}{\int d\Omega}\\
=\frac{1}{3}(v_{c}r_l)^{2}\sin^{2}\theta_{l}
\label{eqvarjz}
\end{multline}
where $r_l$ is the original location of the convective cell,
and $\theta_l$ is the positional latitude from the $z$-axis.
Averaging over all possible positions gives
\begin{equation}
\overline{\mathrm{Var}(j_{z})}=
\frac{\int d\varphi_{l} \int d\theta_{l}\sin\theta_{l}\mathrm{Var}\left(j_{z}\left(\theta_{l}\right)\right)}{\int d\Omega_{l}}=
\frac{2}{9}(v_{c}r_{l})^{2},
\label{eqvarjav}
\end{equation}
which is the same for $j_{x}$ and $j_{y}$. Taking just one
component of the angular momentum gives a slight underestimation
for its magnitude, but simplifies the derivation here, and more so
in the next section where we calculate the average angular
momentum of many cells. The average standard deviation for a
single convective element is then
\begin{equation}
\sigma_j \equiv \overline{\sigma\left(j_z\right)}=\frac{\sqrt{2}}{3}v_{c}r_{l},
\label{eqstdj}
\end{equation}
where $v_{c}\left(r_{l}\right)$ is calculated at the original location of the convective element (cell) $r_{l}$.
The specific angular momentum of a Keplerian orbit at the NS surface is
\begin{equation}
j_\mathrm{NS}=\sqrt{GM_\mathrm{NS}R_\mathrm{NS}},
\label{eqjns}
\end{equation}
so that the ratio between the standard deviation of the
specific angular momentum of a single convective element (cell)
and the minimum required to avoid direct accretion from the equatorial plane is
\begin{multline}
\frac{\sigma_j}{j_\mathrm{NS}} \simeq 0.55
\left(\frac{\mathcal{M}_{c}}{0.1}\right)
\left(\frac{c_{s}}{5000 \kms}\right)
\left(\frac{r_l}{5000 \km}\right) \\
\times
\left(\frac{M_\mathrm{NS}}{1.4M_{\odot}}\right)^{-1/2}
\left(\frac{R_\mathrm{NS}}{25 \km}\right)^{-1/2},
\label{eqjratio}
\end{multline}
where $c_{s}$ is the sound speed given at $r_l$ (the radius of origin of the convective cell),
$\mathcal{M}_c$ is the average convective Mach number,
and typical values for the silicon layer of a pre-collapse core have been inserted.
The choice of $R_\mathrm{NS}\simeq 25 \km$ is due to the protoneutron star (PNS)
needing to cool down before shrinking to estimated radii of observed neutron stars.

\begin{figure}
   \centering
    \includegraphics*[scale=0.54]{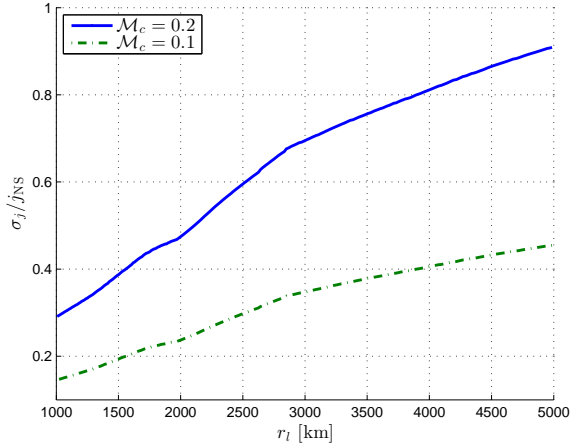} \\
      \caption{Ratio between the standard deviation of the
      specific angular momentum of a single convective mass element
      to the specific angular momentum of a Keplerian orbit around a NS with a radius of
      $R_\mathrm{NS}=25 \km$
      (as given in Equation (\ref{eqjratio})),
      as function of original radius of in-falling material.
      The standard deviation is calculated,
      for our $M_\mathrm{ZAMS}=15M_{\odot}$ stellar model,
      using the local
      sound speed $c_{s}\left(r_{l}\right)$ and for two
      different Mach numbers (given in the inset)
      for the convective velocity at the layer of origin of the convective element ($r_{l}$).
      The values close to unity of this ratio imply that some mass elements can form a temporary accretion disk around the newly formed NS.}
      \label{fig:j1elem}
\end{figure}

We apply Equation (\ref{eqjratio}) to a stellar model
of $M_\mathrm{ZAMS}=15M_{\odot}$
that we evolve with MESA just to the point of core collapse.
Figure \ref{fig:j1elem} shows that the stochastic deviations of specific angular momentum
are close to that of a Keplerian orbit at the NS surface.
This means that some fraction of the in-falling material has
sufficient specific angular momentum to temporarily
form accretion disks around the NSs.

\subsection{Geometrically thick accretion disk}
\label{subsec:thickbelt}

\begin{figure}
   \centering
    \includegraphics*[scale=0.4]{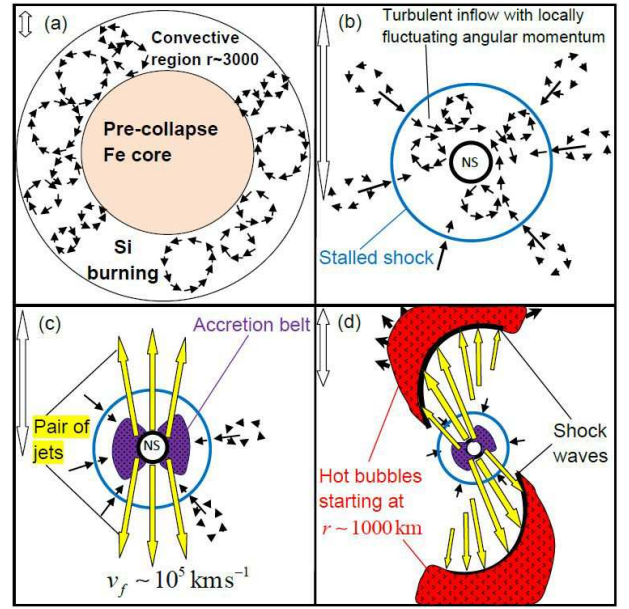} \\
      \caption{A schematic presentation of the proposed scenario. The panels are not exactly to scale, but crudely the two-sided arrow on the upper left of each panel is approximately $500 \km$. The four panels span an evolution time of several seconds.
(a) In the silicon burning shell of the pre-collapse core there is a convective region, at about thousands of km from the center. The convective vortices are the source of the stochastic angular momentum.   
(b) After collapse and the formation of a neutron star (NS) the rest of the in-falling gas passes through the stalled shock. The stochastic spatial distribution of angular momentum in the silicon burning shell is carried inward into the post-shock region.   
(c) The accreted angular momentum changes stochastically in magnitude and direction. For short periods of times, tens of milliseconds, the accreted gas near the NS possesses a net angular momentum. Accretion along and near the temporary poles of the angular momentum axis is inhibited, and a temporary accretion belt is formed around the newly born NS. If the belt exists for a long enough time, several dynamical time, or $> 0.01 \s$, it can spread in the radial direction to form an accretion disk. The belt or disk are assumed to launch two opposite jets with initial velocities of $v_f \approx 10^5 \km \s^{-1}$ (about the escape velocity from the newly formed NS).   
(d) The jets that are launched in varying directions, called jittering jets, penetrate through the gas close to the center, and their shocked gas inflate hot bubbles (see \citealt{Papish2011}). These bubbles expand and explode the star in the jittering jets model \citep{PapishSoker2014a, PapishSoker2014b}.}
      \label{fig:schem}
\end{figure}

The above derivation is limited to the case of a thin accretion disk -
an accretion disk with an opening angle (where there is no gas)
from the angular momentum axis of $\theta=90^\circ$.
The inflowing gas is in the equatorial plane, i.e.,
at latitude of $\theta=90^\circ$ to the angular momentum axis.
If the geometrically thick accretion disk is very close to the NS,
specifically, if $H > r_\mathrm{e}$ and $r_\mathrm{e}$ is not much larger than the radius of the newly formed NS,
we can term it an accretion belt.
Figure \ref{fig:schem} describes schematically our proposed scenario.

The case of an accretion belt may arise when the intermittent accretion disk has no time to spread outward.
Estimating the viscosity as $\nu \approx \alpha c_s H$ 
where $\alpha$ is the viscosity parameter,
and taking for the case of a belt $H \approx r_\mathrm{e}$, the ratio between the viscous and Keplerian timescales is
\begin{multline}
\frac{t_{\nu}}{t_\mathrm{Kep}} \approx 1.4
\left(\frac{\alpha}{0.1}\right)^{-1}
\left(\frac{M_\mathrm{NS}}{1.4M_{\odot}}\right)^{1/2} \\ \times
\left(\frac{R_\mathrm{NS}}{25 \km}\right)^{-1/2}
\left(\frac{c_{s}}{10^5 \km~s^{-1}}\right)^{-1}.
\label{eqtimescales}
\end{multline}
The viscous timescale, in which the disk spreads,
may be between one to a few Keplerian orbit times.
In principle the belt has time in a few dynamical timescales to spread,
first to a geometrically thick and then thin disk.
However, in most cases further mass will be accreted,
mostly with angular momentum in a different direction.
Therefore, while the viscous effect is to form a more well defined accretion disk,
further accreted gas acts mainly to reduce the specific angular momentum and change its direction.
We expect that a belt, or a geometrically thick disk,
will be maintained but with a varying angular momentum axis.
In some cases the belt will be completely destroyed. 
A new one will be formed shortly afterward,
depending on the fluctuations of the specific angular momentum of the accreted gas.

For a thick accretion disk (or a belt) with an opening angle $\theta$
(i.e., the surface of the disk is at an angle $\theta$ from the angular momentum axis,
and the other side at an angle $\theta$ from the opposite direction of the axis),
the inflowing material on the surface of the disk needs a minimum specific angular momentum of
\begin{equation}
j_\mathrm{NS}\left(\theta\right)=\sqrt{GM_\mathrm{NS}R_\mathrm{NS}\sin\theta},
\label{eqjnstheta}
\end{equation}
in order to spiral around the NS surface. It must lose some
angular momentum before being accreted; this specific angular
momentum is only $\sqrt{\sin {\theta}}$ times that required for a
thin disk. From Equation (\ref{eqjnstheta}) we can estimate the
probability for an inflowing parcel of gas to be limited to an
angle, from the angular momentum axis, larger than $\theta_a$. As
our assumptions constrain the specific angular momentum of the
convective elements to $ - v_{c} r_{l} \leq j_{z} \leq v_{c}
r_{l}$, a beta distribution is appropriate,
\begin{equation}
f\left(j_z\right)=
\frac{\left(\frac{1}{2}+\frac{1}{2}\frac{j_z}{v_{c}r_{l}}\right)^{\alpha-1}
\left(\frac{1}{2}-\frac{1}{2}\frac{j_z}{v_{c}r_{l}}\right)^{\beta-1}}{B\left(\alpha,\beta\right)},
\label{eqjpdf}
\end{equation}
where $f(j_z)$ is the probability density function for a convective element to have a specific angular momentum component $j_z$,
$\alpha$ and $\beta$ are shape parameters determined by the expectation value and variance,
and $B\left(\alpha,\beta\right)$ is the beta function.
An expectation value of zero for $j_z$ and the variance from Equation (\ref{eqvarjav}) give $\alpha=\beta=7/4$.
The desired probability function is given by
\begin{multline}
\xi\left(\theta_a\right)=
2\int_{j_z= \min \left(j_\mathrm{NS}\sin\theta_a, v_{c}r_{l}\right)}^{j_z=v_{c}r_{l}}dj_z f\left(j_z\right) = \\
2\left(1 - I_x\left(\alpha,\beta\right)\right),
\label{eqxi}
\end{multline}
where $\xi\left(\theta_a\right)$ is the probability that
accretion of a convective parcel of gas will be limited to an angle of
$\pi - \theta_a > \theta> \theta_a$, and
$I_x\left(\alpha,\beta\right)$
is the regularized incomplete beta function with
\begin{equation}
x=\min\left(\frac{1}{2}+\frac{1}{2} \frac{j_\mathrm{NS}\sin\theta}{v_{c}r_{l}} ,1\right).
\label{eqxbeta}
\end{equation}
The factor 2 in front of the integral (possible with the $\alpha=\beta$ symmetry)
is for the two sides of the equatorial plane:
between $\theta_a$ and $90^\circ$, and between $90^\circ$ and $(180^\circ-\theta_a)$.

\begin{figure}
   \centering
 \begin{tabular}{cc}
{\includegraphics*[scale=0.52]{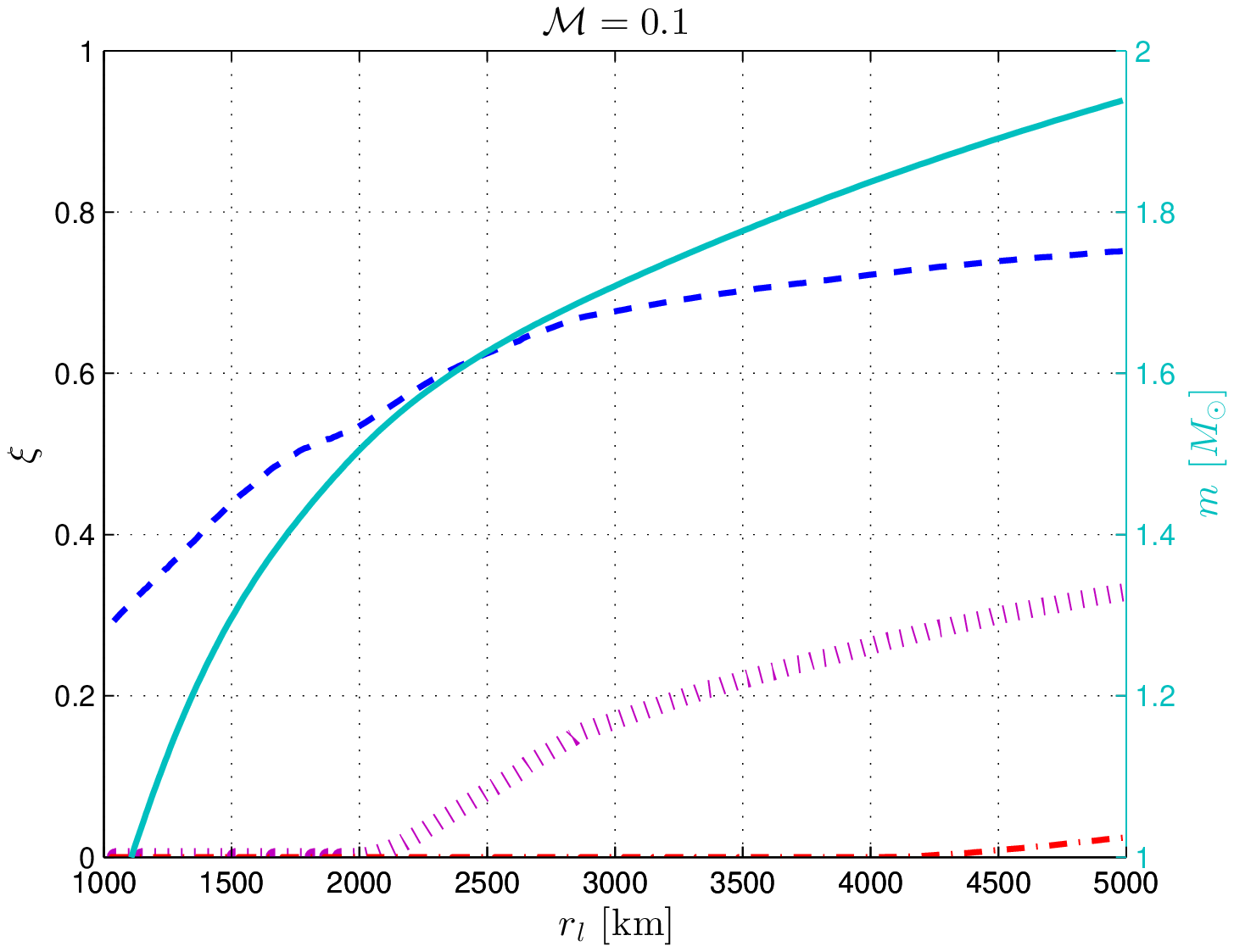}} \\
{\includegraphics*[scale=0.52]{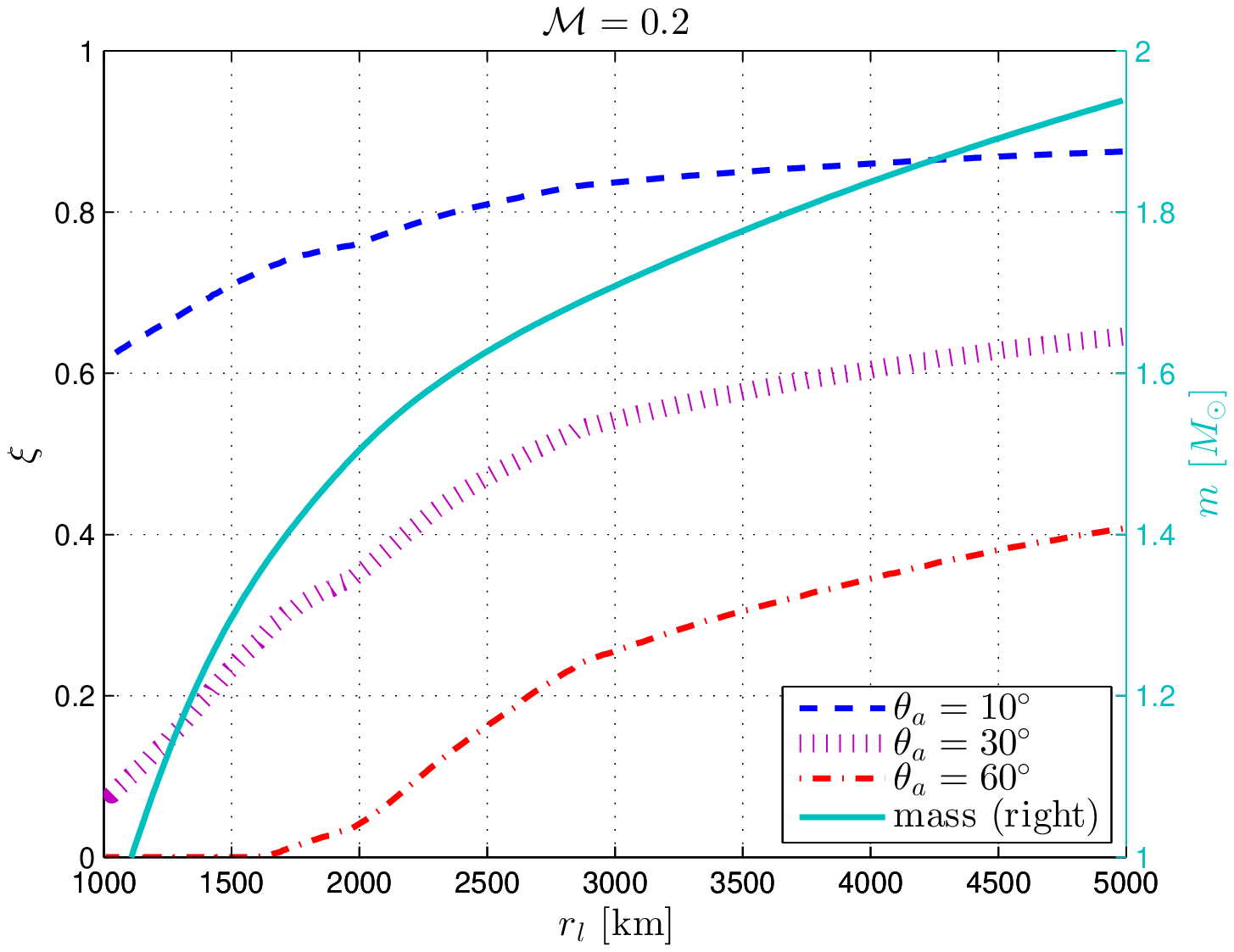}} \\
\end{tabular}
      \caption{
      The probability for accretion onto the NS to be limited to an angle $\theta > \theta_a$
      as a function of radius of origin of the convective mass
      element in the core, and for three values of the angle from the angular momentum axis $\theta_a$ (given in the inset).
      This is calculated
      for convective Mach numbers of $\mathcal{M}_c=0.1$ (top panel)
      and $\mathcal{M}_c=0.2$ (bottom panel),
      and assuming a beta distribution (Equation (\ref{eqjpdf})).
      For example, for a convective Mach number of $\mathcal{M}_c=0.2$ (bottom panel),
      material falling from $r_{l}=4000 \km$ has a $60 \%$ probability of having
      a specific angular momentum limiting the accretion to take place from an angle larger than $30^\circ$
      from the angular momentum axis. This probability is $26 \%$ for the Mach=0.1 case.
      Limitation to $\theta > \theta_a=0$ (all lines on the $\xi=0$)
      would imply a spherical accretion flow, since all angles are allowed --
      this is the case when there is no stochastic motion at the shell of origin of the inflowing gas.
      On the other hand, a value of $\xi=1$ for all angles $\theta_a$ implies
      that the gas has a specific angular momentum too high to be directly accreted onto the NS.
      A thin accretion disk will be formed, and accretion will proceed by angular momentum loss in the disk.
      Values of $\xi>0$ lead to the formation of a low density accretion funnel along the angular momentum axis.
      The solid line shows the inwards baryonic mass at each radius of the pre-collapse core (refers to the right axis).}
      \label{fig:theta1elem}
\end{figure}

This probability for a given angle $\theta_a$ as function
of the radius of origin $r_l$ can be calculated from Equations
(\ref{eqxi}) and (\ref{eqxbeta}) for a given limiting angle $\theta_a$.
We present this in Fig. \ref{fig:theta1elem}
for the same stellar model used for Fig. \ref{fig:j1elem}.
To better understand the meaning of $\xi(\theta_a)$ we can examine limiting cases.
If there is no stochastic angular momentum at all, i.e., $v_{c}r_{l} \ll j_{NS}$,
then $\xi=2\left(1 - I_1\left(\alpha,\beta\right)\right)=0$ for all angles.
Namely, the probability for limiting the angle is zero
as expected since there is no angular momentum and hence
each parcel of gas can in principle be accreted from any direction.
If the angular momentum fluctuations are huge, $v_{c}r_{l} \gg j_\mathrm{NS}$,
then $\xi=2\left(1 - I_{1/2}\left(\alpha,\beta\right)\right)=1$.
Namely, for all angles the probability for accretion above the angle is 1, and hence below the angle is zero.
This is true even for $\theta=90^\circ$,
which implies that the accreted gas is stopped from inflowing
due to a centrifugal barrier at radii larger than the NS radius.
Further angular momentum loss in a viscous disk will allow accretion.
Other examples are in the caption of Fig. \ref{fig:theta1elem}.

\section{ACCRETION OF MULTIPLE CONVECTIVE ELEMENTS}
\label{sec:multielem}

The accretion of material from a single convective element is a
simplified case, as in reality many elements with close radii of
origin may undergo accretion at overlapping times. For
simultaneous accretion of multiple convective elements (which we
assume to have equal masses), with stochastically varying
velocities, the variance of the specific angular momentum becomes
\begin{equation}
\mathrm{Var}(j_{z,N})=\frac{2}{9N}(\mathcal{M}_{c}c_{s}r_{l})^{2},
\label{eqvarjn}
\end{equation}
where $N$ is the number of convective elements in the shell from which the mass is accreted at the given specific time,
and $\mathcal{M}_c$ is the average Mach number of the convective cells.
This is Equation (\ref{eqvarjav}) with a factor of $N^{-1}$ and with $\mathcal{M}_{c} c_{s}=v_{c}$.
We get a narrower distribution of the angular momentum for $N>1$;
for the beta distribution we assumed in the previous section we get
\begin{equation}
\alpha=\beta=\frac{9N}{4}-\frac{1}{2}.
\label{eqalphabetan}
\end{equation}
The number $N$ can be estimated using the mixing-length
(which is proportional to the pressure scale-height)
or from the relevant mode order,
such as those used by \cite{CouchOtt2013,CouchOtt2014} or \cite{MuellerJanka2014b}.
For example, \cite{CouchOtt2013}
use sinusoidal perturbations (their Equation (1))
which directly give us the number of elements in a shell.
For example, their equation (1) would give for $n=3$ and $n=5$ modes
a total of $N=12$ and $N=40$, respectively.
A rough estimation using MLT,
taking spherical elements with size $r_\mathrm{elem}=\alpha_\mathrm{MLT}H_P$
($\alpha_\mathrm{MLT}$ being the MLT parameter and $H_P$ the pressure scale height)
so that
\begin{equation}
N \approx \frac{4\pi r_l^2}{\pi r_\mathrm{elem}^2},
\label{eqnmlt}
\end{equation}
gives values of $N \approx 20-40$ for the region of interest.

As we consider non-rotating stars,
the total angular momentum is zero.
We consider fluctuations in angular momentum,
not only within a shell, but also between shells.
When a shell has (temporarily) non-zero angular momentum,
other shells will compensate with angular momentum axes with other orientations,
for a total of zero angular momentum.
Over time an exchange of angular momentum takes place between shells
(as well as between convective elements).
Even for rotating stars,
convective regions may give rise to temporary deviations from the angular momentum dictated by the rotation
(hence leading to jittering jets).
It is important to note that
\cite{CouchOtt2013,CouchOtt2014} and \cite{MuellerJanka2014b}
considered only fluctuations within each shell,
but the sum of angular momentum was zero in each shell.
This might explain why they did not get accretion belts.

\begin{figure}
   \centering
 \begin{tabular}{cc}
{\includegraphics*[scale=0.52]{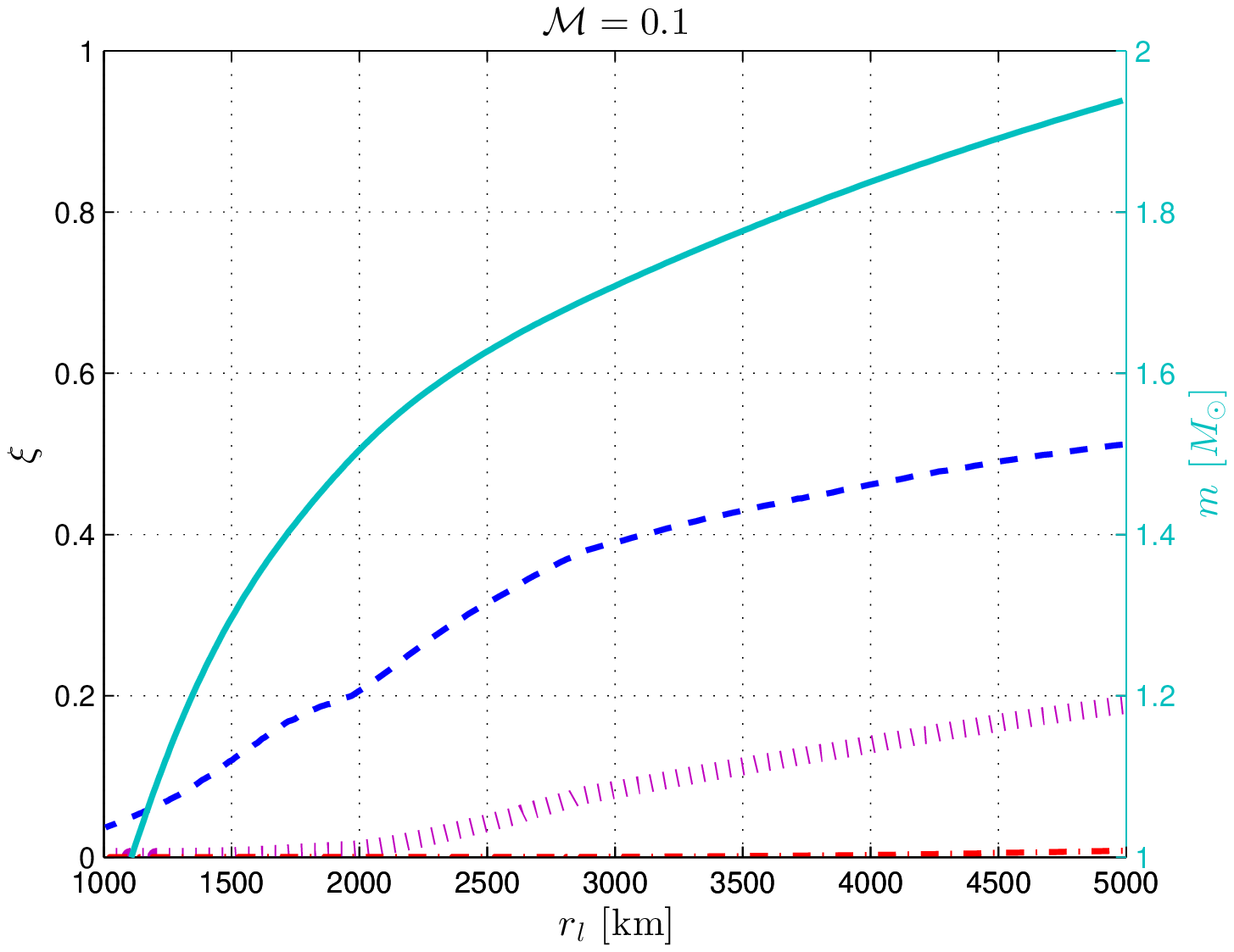}} \\
{\includegraphics*[scale=0.52]{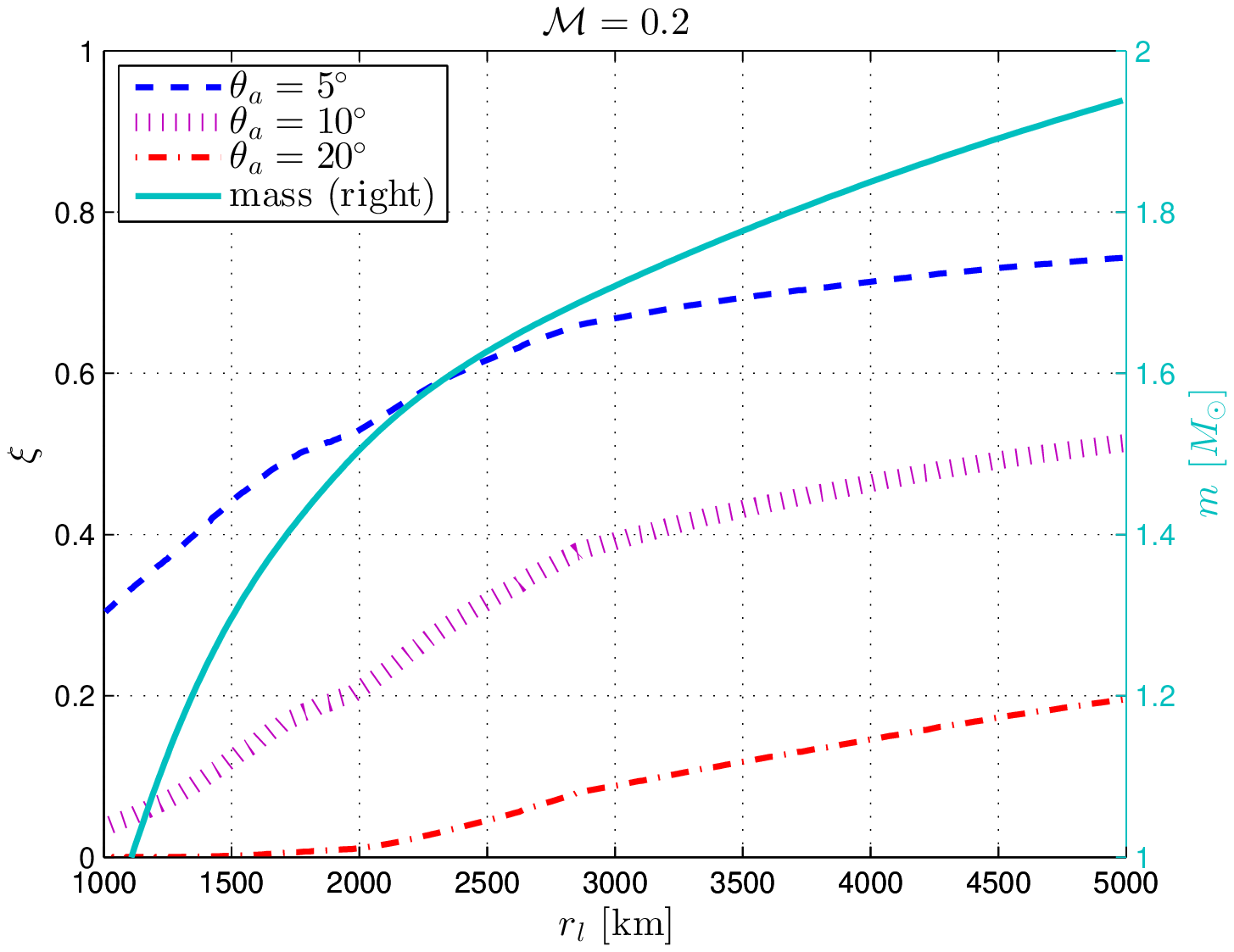}} \\
\end{tabular}
      \caption{Same as Fig. \ref{fig:theta1elem} but for accretion of a shell with an $n=3$ mode
      (shell with $N=12$ ``blobs'') using Equation (1) of \cite{CouchOtt2013}, and for different limiting angles.
      For example, for a convective Mach number of $\mathcal{M}_c=0.2$ (bottom panel),
      about $40\%$ of the accreted mass from a shell starting at $r_l=3000 \km$
      will not be allowed to be accreted within $10^\circ$ from the angular momentum axis
      (intersection of the the blue dashed line with the $r_{l}=3000 \km$ line),
      and $9\%$ of the mass starting at the same shell will not be able to be accreted from within $20^\circ$
      from the angular momentum axis (intersection of the green dotted line with the $r_{l}=3000 \km$ line).
      These probabilities are $9 \%$ and $0.25 \%$ for a
      convective Mach number of $\mathcal{M}_c=0.1$, respectively.
      The implication is that a low density funnel is formed along the two opposite sides of the axis.}
      \label{fig:thetan3}
\end{figure}

As shown in Fig. \ref{fig:thetan3},
the inclusion of several convective elements results in smaller opening angles (thicker disks)
for the thick disk than in the single parcel presented in Fig. \ref{fig:theta1elem}.
As the convective region is turbulent and disorderly,
this picture of symmetric accretion is not an accurate description.
The accretion process will be something between the parameters of
Fig. \ref{fig:theta1elem} and those in Fig. \ref{fig:thetan3}.

The meaning of figure \ref{fig:thetan3} is as follows.
If there was no turbulence at all,
then all lines would be on the $\xi=0$ axis,
implying that angular momentum does not prevent any gas from being accreted from all angles.
However, the turbulence and the resulting stochastic angular momentum of the accreted mass imply that
a substantial fraction of the mass is prevented from being accreted from a direction close to the angular momentum axis
(on both opposite sides of the angular momentum axis).
Figure \ref{fig:thetan3} shows, for example, that assuming an average convective Mach number of $\mathcal{M}_c=0.2$,
for mass originating at $r_l=2000 \km$ on average $20 \%$ of the mass cannot be accreted at all from within an angle of $10^\circ$
from the angular momentum axis because of a centrifugal barrier.
The outcome is the formation of two low-density opposite funnels of the in-falling gas along the angular momentum axis.
Due to the stochastic nature,
the angular momentum axis is not a constant axis,
but rather varies with time; i.e., it jitters.

\section{DISCUSSION AND SUMMARY}
\label{sec:summary}

We have studied the formation of intermittent accretion disks by
the collapsing convective regions of the core in CCSNe.
This study is motivated by the usage of
convective core regions, which might be more vigorous than what
the MLT gives (e.g., \citealt{Arnett2009},
\citealt{Viallet2013}), to facilitate the revival of the stalled
shock \citep{CouchOtt2013, CouchOtt2014, MuellerJanka2014b}.
It is noteworthy that even with the additional turbulent pressure,
these aforementioned works employed either an enhancement factor for
the neutrino heating \citep{CouchOtt2014},
or imposed axisymmetry \citep{MuellerJanka2014b},
which has already been shown to ease shock revival (e.g. \citealt{Couch2013}).
Namely, it is not entirely clear that the insertion of strong turbulence
will revive the shock in neutrino based mechanisms.

Even if the stalled shock in CCSNe is revived, the desired explosion
energy of $E_{\rm exp} \ga 10^{51} \erg$ is unlikely to be
achieved \citep{Papishetal2015}. The convective regions on the
other hand are likely to lead to the formation of intermittent
accretion disks that can launch jets \citep{GilkisSoker2014} that
are more likely to explode the star than the revival of the
stalled shock \citep{PapishSoker2014a,PapishSoker2014b}.
In this model the jets are presumably launched by the
intermittent accretion disks formed around the newly born NS.
By jets we refer to collimated outflows from the innermost region of the accretion disk,
rather than wide disk winds launched from the shallower potential well of the extended disk region
(a less energetically efficient case).
In that respect disk-winds can also do the job in the jittering jets model,
as long as they are sufficiently collimated (a half opening angle of $\la 20^\circ$)
for them to penetrate out to about $1000 \km$ or more \citep{PapishSoker2014a}.
However, the
jet properties are not as those of jets formed by preset
magnetic fields and core rotation,
as in the simulations by \cite{Mostaetal2014}.
Therefore, the failure of the jets simulated by \cite{Mostaetal2014}
to explode the star does not imply the failure of the jittering jets model.

We here extended our earlier study (\citealt{GilkisSoker2014})
in discussing the formation of a thick accretion disk (or an
accretion belt), and not only a thin accretion disk, and in
referring specifically to the convection topology used by
\cite{CouchOtt2013}, \cite{CouchOtt2014}, and
\cite{MuellerJanka2014b}.
The ordered structure introduced in these previous works
(with zero angular momentum in each shell)
is perhaps not representative of the turbulent flow structure.
If that is the case, it
overlooks the possibility of angular momentum deviations between shells.
We considered fluctuations of angular momentum between shells,
not only within shells,
and found that these between-shells angular momentum fluctuations
can lead to intermittent thick accretion disk (belt) formation.

As evident from Fig. \ref{fig:thetan3}
the accretion from such a convective region forms an accretion
belt (or a thick accretion disk) that leaves a funnel along the
polar directions. The general accretion flow not only leaves two
opposite funnels of a very low density, but around the funnels the
gas is rapidly rotating. The turbulent accretion belt is very
likely to amplify magnetic fields. This is similar to the finding
of \cite{Masadaetal2014} of the development of turbulence and
magnetic field amplification around a nascent proto-NS. The
funnel, rotation, and magnetic field amplification are the
ingredients that very likely form jets. The formation of such jets
is along the jittering-jets scenario \citep{PapishSoker2014a,
PapishSoker2014b}.

There are some difficulties in relating the stochastic angular
momentum, clearly present in the convective region prior to
collapse, to the angular momentum of material reaching the
newly formed compact object. This is because the accreted gas goes
through the shock wave moving through the turbulent region from
the shock down to the newly born NS or black hole
(BH). The turbulent region between the stalled shock and
the newly born NS or BH might increase or decrease the variance of
the specific angular momentum. Multidimensional hydrodynamic
simulations are required to resolve this question.
We note that \cite{CouchOtt2013,CouchOtt2014} and \cite{MuellerJanka2014b}
introduced vortices in a way that the total angular momentum in each shell sums-up to zero.
In our study, the formation of an accretion disk results from angular momentum fluctuations between shells in the pre-collapse core.
This is one of the reasons the aforementioned studies could not obtain an accretion disk.
Another reason is that the shear between the newly formed NS and the accreted gas must be calculated with high resolution.
We expect that appropriate initial perturbations and high resolution near the NS will result in accretion belt formation.

We here try to estimate the stochastic specific angular momentum from existing simulations.
Recent studies (e.g., \citealt{CouchOtt2014}) have focused on the
turbulent energy in the gain region, giving typical values for
the mass $M_\mathrm{turb}$ and energy $E_\mathrm{turb}$ in this turbulent region.
We take
\begin{equation}
\overline{v_\mathrm{turb}} \equiv \sqrt{\frac{2E_\mathrm{turb}}{M_\mathrm{turb}}},
\label{eqvturb}
\end{equation}
use Equation (\ref{eqstdj}) to approximate
$j_\mathrm{turb} \approx \sigma_j = \frac{\sqrt{2}}{3} v_\mathrm{turb} r_{l}$,
and substitute typical values to derive
\begin{multline}
\frac{j_\mathrm{turb}}{j_\mathrm{NS}} \approx 0.2
\left(\frac{E_\mathrm{turb}}{2 \times 10^{49} \erg}\right)^{1/2}
\left(\frac{M_\mathrm{turb}}{0.05 M_{\odot}}\right)^{-1/2} \\
\times
\left(\frac{r_{l}}{150 \km}\right)
\left(\frac{M_\mathrm{NS}}{1.4M_{\odot}}\right)^{-1/2}
\left(\frac{R_\mathrm{NS}}{25 \km}\right)^{-1/2}.
\label{eqjturb}
\end{multline}
This is valid for symmetric as well as turbulent initial
conditions, as turbulence around the NS arises either way
(e.g., \citealt{MuellerJanka2014b}).
Comparing Equation (\ref{eqjturb}) with Equation (\ref{eqjratio})
with the aid of Equation (\ref{eqvarjn}), we find that the case
given by Equation (\ref{eqjturb}) corresponds to a total number of
convective cells in the accreted layer of $N \approx 7 -8$. This
will give funnels similar (and even somewhat larger) than those
depicted in figure \ref{fig:thetan3} which is given for $N=12$.
This crude estimate suggests that the passage of the material
through the stalled shock does not reduce much the stochastic
behavior of the angular momentum.

We summarize by restating our main finding that the pre-collapse
turbulence structures introduced by
\cite{CouchOtt2013,CouchOtt2014} and \cite{MuellerJanka2014b}
might lead to the formation of intermittent accretion disks.
These in turn are likely to launch jets that play a much more important
role in exploding the the star than the extra pressure of the
turbulent motion on the stalled shock region.

We thank an anonymous referee for comments that improved the presentation of our results.
This research was supported by the Asher Fund for Space Research
at the Technion.


\label{lastpage}

\end{document}